\documentclass[nofootinbib,aps,prl,twocolumn,superscriptaddress,groupedaddress]{revtex4} 
\usepackage{graphicx}  % needed for figures
\usepackage{dcolumn}   % needed for some tables
\usepackage{bm}        % for math
\usepackage{amssymb}   % for math
\usepackage{amsmath}
\usepackage[dvipsnames]{xcolor}
\usepackage{footmisc}
\usepackage{color}
\usepackage{multirow}
\usepackage{footmisc}
\begin{document}

\title{The Effect of Intra-Layer Bonding on Electron-Optical Phase~Images of~Few-Layer~WSe$_2$}

% remove these 3 lines before journal submittal.
% \centerline{author list dated 24 March 2010}
% end removal before journal submittal
%
\affiliation{Peter Gr\"unberg Institute 9 (PGI-9), Forschungszentrum J\"ulich, D-52425 J\"ulich, Germany}
\affiliation{Ernst Ruska-Centre for Microscopy and Spectroscopy with Electrons (ER-C), Forschungszentrum J\"ulich, D-52425 J\"ulich, Germany}
\affiliation{Peter Gr\"unberg Institute 5 (PGI-5), Forschungszentrum J\"ulich, D-52425 J\"ulich, Germany}
\affiliation{Peter Gr\"unberg Institute 1 (PGI-1) and Institute for Advanced Simulations (IAS-1), Forschungszentrum J\"ulich, D-52425 J\"ulich, Germany}
\affiliation{Institute for Theoretical Solid State Physics, RWTH Aachen University, D-52056 Aachen, Germany}
\affiliation{Physique des mat\'eriaux et nanostructures (NanoMat), CESAM and D\'epartement de Physique, Universit\'e de Li\`ege (B5), B-4000 Li\`ege, Belgium}
\affiliation{European Theoretical Spectroscopy Facility, http://www.etsf.eu}
\affiliation{Gemeinschaftslabor f\"ur Elektronenmikroskopie (GFE), RWTH Aachen University, D-52074 Aachen, Germany}
\author{S. Borghardt} \affiliation{Peter Gr\"unberg Institute 9 (PGI-9), Forschungszentrum J\"ulich, D-52425 J\"ulich, Germany}
\author{F. Winkler} \affiliation{Ernst Ruska-Centre for Microscopy and Spectroscopy with Electrons (ER-C), Forschungszentrum J\"ulich, D-52425 J\"ulich, Germany} \affiliation{Peter Gr\"unberg Institute 5 (PGI-5), Forschungszentrum J\"ulich, D-52425 J\"ulich, Germany}
\author{Z. Zanolli} \affiliation{Peter Gr\"unberg Institute 1 (PGI-1) and Institute for Advanced Simulations (IAS-1), Forschungszentrum J\"ulich, D-52425 J\"ulich, Germany} \affiliation{Institute for Theoretical Solid State Physics, RWTH Aachen University, D-52056 Aachen, Germany} \affiliation{European Theoretical Spectroscopy Facility, http://www.etsf.eu}
\author{M.~J. Verstraete} \affiliation{Physique des mat\'eriaux et nanostructures (NanoMat), CESAM and D\'epartement de Physique, Universit\'e de Li\`ege (B5), B-4000 Li\`ege, Belgium} \affiliation{European Theoretical Spectroscopy Facility, http://www.etsf.eu}
\author{J. Barthel} \affiliation{Ernst Ruska-Centre for Microscopy and Spectroscopy with Electrons (ER-C), Forschungszentrum J\"ulich, D-52425 J\"ulich, Germany} \affiliation{Gemeinschaftslabor f\"ur Elektronenmikroskopie (GFE), RWTH Aachen University, D-52074 Aachen, Germany}
\author{R.~E. Dunin-Borkowski}  \affiliation{Ernst Ruska-Centre for Microscopy and Spectroscopy with Electrons (ER-C), Forschungszentrum J\"ulich, D-52425 J\"ulich, Germany} \affiliation{Peter Gr\"unberg Institute 5 (PGI-5), Forschungszentrum J\"ulich, D-52425 J\"ulich, Germany}
\author{B. Kardynal} \affiliation{Peter Gr\"unberg Institute 9 (PGI-9), Forschungszentrum J\"ulich, D-52425 J\"ulich, Germany}
%
% list_of_visitor_addresses_r2.tex            24 March 2010
%  available symbols are:
%  $\ast, \dag, \ddag, \S, \P, $\|$, $\ast\ast$, \dag\dag, \ddag\ddag ,\#
%
% \collaboration{The D0 Collaboration\footnote{with visitors from
% $^{a}$  %{alton}
% Augustana College, Sioux Falls, SD, USA,
% $^{b}$  %{burdin}
% The University of Liverpool, Liverpool, UK,
% $^{c}$  %{haas,partridge}
% SLAC, Menlo Park, CA, USA,
% $^{d}$  %{juste}
% ICREA/IFAE, Barcelona, Spain,
% $^{e}$  %{luna-garcia}
% Centro de Investigacion en Computacion - IPN,
%   Mexico City, Mexico,
% $^{f}$  %{podesta-lerma}
% ECFM, Universidad Autonoma de Sinaloa, Culiac\'an, Mexico,
% $^{g}$  %{weber}
% and Universit{\"a}t Bern, Bern, Switzerland.
% %$^{?}$  %{hooper}
% %Visitor from Bradley University, Peoria, IL, USA.
% %$^{?}$  %{kozminski}
% %Visitor from Lewis University, Romeoville, IL, USA.
% %$^{\ddag}$  %{deceased}
% %Deceased.
% }} \noaffiliation
%
\vskip 0.25cm

\begin{abstract}
The quantitative analysis of electron-optical phase images recorded using off-axis electron holography often relies on the use of computer simulations
of electron propagation through a sample. 
However, simulations that make use of the independent atom approximation are known to overestimate experimental phase shifts by approximately 10\%, as they neglect bonding effects.
Here, we compare experimental and simulated phase images for few-layer WSe$_2$. We show that a combination of pseudopotentials
and all-electron density functional theory calculations can be used to obtain accurate mean electron phases, as well as improved
atomic-resolution spatial distribution of the electron phase. The comparison demonstrates 
a perfect contrast match between experimental and simulated atomic-resolution phase images for a sample of precisely know thickness.
The low computational cost of this approach makes it suitable for the analysis of large electronic systems, including defects, substitutional atoms and material interfaces.
\end{abstract}

\maketitle

The complex wavefunction of electrons that have passed through a sample in the transmission electron microscope (TEM) 
can be reconstructed using the technique of off-axis electron holography.
For a non-magnetic sample, the phase of the electron wavefunction is related to the 
three-dimensional electrostatic potential in the specimen and, in the absence of dynamical scattering, is proportional to the integral of the electrostatic potential in the electron beam direction \cite{dunin-borkowski}.
As a result of the high spatial resolution of TEM, off-axis electron holography is therefore a powerful technique for the characterisation of local variations in electrostatic potential in functional materials at the nanoscale \cite{midgley}. 
 
In general, the conversion of a recorded phase image into a potential is non-trivial and often has to be supported by atomistic computer simulations \cite{cowley}.  
An approach that is used frequently makes use of the independent atom approximation (IAA) and involves representing
the crystal potential as a superposition of electrostatic potentials of individual isolated atoms \cite{kirkland}.
As the effects of bonding are neglected, the results of simulations based on this approximation overestimate the
mean phase of the electron wavefunction when compared to experimental measurements \cite{kruse}. 
The accuracy of calculated mean electron phases has been shown to improve when using 
density functional theory (DFT) for the calculation of electrostatic potentials to take
bonding effects into account \cite{kim,kruse2}. However, to the best of our knowledge, no comparison between the 
DFT-calculated atomic-resolution spatial distribution of the electron phase and high-resolution electron holography 
experiments has been performed. 
The importance of developing a technique suitable for such comparisons is growing,
as the operation of electronic and optoelectronic devices relies more frequently on or is affected by individual atoms
and local structure variations. 
For example, the electrical properties of modern transistors are often determined by single dopant atoms in their channels and
individual nitrogen-vacancy centers in diamond are used for quantum sensors \cite{pierre,cai}. The identification and characterisation
of similar defects are, hence, of great importance for the understanding and further improvement of the performance of both 
these systems and future optoelectronic devices.
This requirement applies especially to the emerging field of two-dimensional materials, where, due to their thickness,
the manipulation of individual atoms
strongly affects the properties of the materials.

In addition to the approximations that are used in simulations, experimental uncertainties often prevent quantitative comparisons with measurements.
The most common experimental uncertainties include a poor knowledge of the sample thickness, which can be difficult to determine with sufficient precision at high spatial resolution, as well as the possible presence of crystal defects, surface damage, reconstructions and contamination \cite{pennington}. 
In this regard, transition metal dichalcogenides (TMDs), which have the chemical composition MX$_2$ where M and X
denote a transition metal and a chalcogen, respectively, are an exception. They are
layered materials that can be prepared with a thickness of an integer number of monolayers and do not form surface 
dangling bonds, which are responsible for surface reconstructions in other materials. Although surface contamination still presents
a major problem for off-axis electron holography of TMDs, clean areas can often be identified and studied \cite{florian}.

The possibility of reconstructing the true crystal potentials of TMDs using off-axis electron holography is highly attractive, both because they provide a model system that allows the phase evolution of electrons passing through thin samples to be understood and because they have 
properties that are promising for new optoelectronic applications, including transistors, light sources and 
photodetectors \cite{kis,ross,lopez}.
Recent advances in the engineering of  TMDs pave the way for atomically sharp lateral and vertical heterostructures of these materials \cite{lee,gong,mahjouri}, while as-yet-unknown defects in WSe$_2$ monolayers have been found to emit single photons \cite{srivastava,he}.

Here, we show that both the mean phase of few-layer TMDs measured using off-axis electron holography
and the atomic-scale spatial redistribution of the phase
can be described accurately using simulations based on three-dimensional potentials that 
include the effects of bonding. In order to demonstrate these benefits, we compare average
electron phases obtained using both DFT and IAA calculations and their spatial distributions with our experimental results \cite{florian}. 
The effects of bonding are assessed by comparing crystal potentials obtained from calculations that include bonding effects with ones that neglect them.

\begin{figure}
\includegraphics{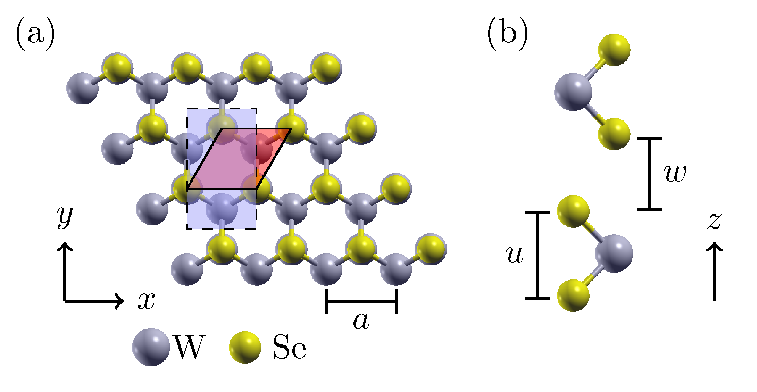}
\caption{\label{fig:geometry}WSe$_2$ lattice geometry. (a) Top view
showing multiple unit cells. The red rhombus marks the unit cell
used in DFT calculations, while the blue rectangle marks the cell used in multislice simulations of electron-optical phase images.
(b) Side view of a unit cell, which comprises a stack of two monolayers.}
\end{figure}

As a preliminary step for the calculation of electrostatic potentials, pseudopotential 
DFT relaxation calculations were carried out in order to obtain the crystal
structure of WSe$_2$, which is shown in Fig.~\ref{fig:geometry}, for thicknesses of up to five layers. 
The pseudopotential calculations were performed  using the ABINIT software package \cite{gonze1,gonze2}
with the Perdew-Zunger-Ceperley-Alder LDA exchange-correlation functional \cite{perdew} and optimized norm-conserving
Vanderbilt pseudopotentials, following the scheme of Hamann \cite{hamann}. 
Tolerances on the maximal force and stress in the relaxation calculations were set to $5\cdot 10^{-5} \frac{\text{Ha}}{a_0}$ 
and $5\cdot 10^{-7} \frac{\text{Ha}}{a_0^3}$, respectively. Convergence 
studies were used to choose a plane wave cut-off energy of $32$~Ha and an 8$\times$8$\times$1 Monkhorst-Pack \textit{k}-point grid \cite{monkhorst}, thereby ensuring
convergence of the system's total energy below 0.1\%. In order to avoid interactions between periodic images of the finite thickness slab,
a vacuum layer with a thickness of $22$~\AA\ was included in the supercell between the outermost Se planes of
two neighbouring periodic images of the slab.
For calculations of electrostatic potentials, the spatial resolution of the DFT calculation was improved by
increasing the plane wave cut-off energy to $48$~Ha. The high spatial resolution was necessary
for the treatment of sharp electrostatic potentials in the core regions.

The plane wave basis sets and pseudopotentials used in this work reduce the number of active electrons 
and the kinetic energy cutoff. This is common practice, and gives access to large system sizes for defects or 
heterostructures, but it does not explicitly give the core electron charge, which is frozen out of the pseudopotential. 
In order to access the full electrostatic potential, we apply a correction scheme, as reported in \cite{wang}.
For each element of interest (Se and W), isolated-atom $\Gamma$-point all-electron  calculations in a cubic supercell with a side length of
$10.5$~\AA\ were performed using the Elk code with its default computation parameters \cite{elk}. The resulting 
electrostatic potentials were compared with individual-atom pseudopotential calculations performed under identical conditions.
The difference, which results from the modified core potentials in the pseudopotential method, was saved for further use. 
The complete crystal unit cells were then treated using 
pseudopotential calculations and the pre-calculated difference terms were added in order to obtain full electrostatic potentials for subsequent off-axis electron holography simulations.

Two different methods that both neglect the effects of bonding were compared with the DFT results. First,
crystal potentials were determined by the IAA method using elastic electron scattering factors from
the literature \cite{weickenmeier}. Second, independent DFT (IDFT) calculations were used to obtain the 
electrostatic potentials. In the latter case, isolated-atom pseudopotential calculations were 
performed for individual atoms and their electrostatic potentials were then superimposed 
to obtain crystal potentials. These potentials were also corrected using the all-electron terms, as described
for the DFT method. One purpose of using the IDFT method was to determine the change in spatial electron density resulting from bonding effects by evaluating the difference between spatial electron densities obtained by the IDFT and DFT methods.
The second purpose of using the IDFT method was to rule out numerical issues as the reason for any differences between the DFT and IAA calculations by matching the numerical parameters used for the DFT and IDFT methods.

\begin{figure}
\includegraphics{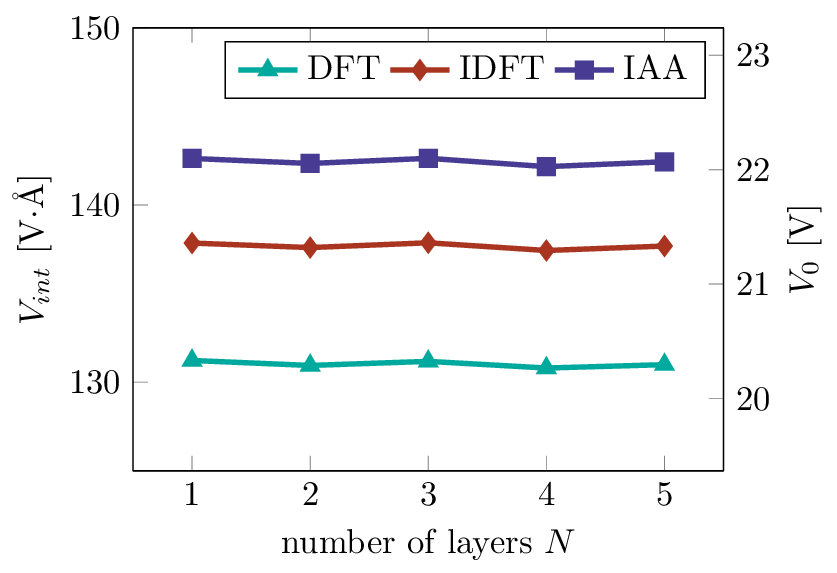}
\caption{\label{fig:mip}Integrated plane-averaged electrostatic potential per layer (left axis)
as well as mean inner potential (right axis)
for few-layer WSe$_2$ calculated using three
different methods. The lowest integrated potential per layer is found when using the DFT method.
The IAA and IDFT methods, which do not take bonding effects into account, yield higher integrated potentials.
}
\end{figure}

In order to compare spatially averaged electrostatic potentials obtained using the three different methods, they were averaged within the $x$-$y$ plane, integrated in the $z$-direction and normalized to the number of layers $N$, according to the expression

\begin{equation*}
 V_{int} = \frac{1}{N}\frac{\int V(x,y,z) dx dy dz}{\int dxdy} .
\end{equation*}

\noindent The integrated plane-averaged potential, $V_{int}$, can be related approximately to the more commonly
used mean inner potential, $V_0$, by dividing $V_{int}$ by the bulk layer periodicity, which
is not defined for few-layer systems.
Figure~\ref{fig:mip} shows both $V_{int}$ and $V_0$
plotted as a function of the number of layers.
The results obtained using all three methods are found to be independent of the number of layers, 
as a result of the weak inter-layer interactions in the material and the absence of surface effects. Furthermore, the DFT
method yields the lowest potentials, whereas the IAA and IDFT results exceed the DFT values by
approximately 9\% and 6\%, respectively. The 
difference between the IDFT and IAA values can be explained by the use of a different computational technique: the scattering
factors that were used as input for IAA electrostatic potentials were obtained from relativistic Hartree-Fock calculations \cite{kruse2}.

\begin{figure}
\includegraphics{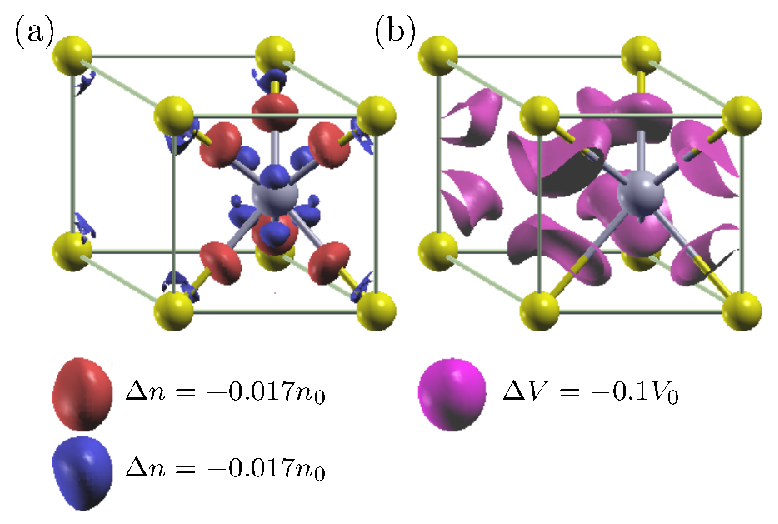}
\caption{\label{fig:isos}(a)~Change in spatial electron density and (b)~change in electrostatic potential 
in a WSe$_2$ monolayer unit cell between the DFT and IDFT methods. 
(a) shows isosurfaces in the difference in spatial electron density corresponding to
$\pm 0.017n_0$, where $n_0$ is the bulk average electron density in the material. A positive value denotes a higher electron density in the DFT method.
(b) shows isosurfaces in the difference in electrostatic potential corresponding to $-0.1V_0$, where $V_0$ is the bulk mean inner potential of  WSe$_2$ \cite{footnote}.
The negative value denotes a lower electrostatic potential in the DFT method.}
\end{figure}

In order to relate differences between the DFT and IDFT calculations to a spatial change in electron 
density associated with bonding, differences between both the spatial electron densities and the
electrostatic potentials were calculated for a WSe$_2$ monolayer, as shown in
Fig.~\ref{fig:isos}. In comparison to IDFT, DFT shows a higher electron density in the interstitial 
regions of the crystal and a correspondingly lower electron density in the proximity of the nuclei. This shift in
electron density is to be expected for covalent bonding between W and Se atoms. As a consequence of the shift 
in electron density, the electrostatic potential is decreased along the bonding directions close to the Se nuclei and within the Se columns.

Figure~\ref{fig:isos} shows that the relationship between $n$ and $V$ is non-linear, with a small shift in electron density 
leading to a significant decrease in electrostatic potential.
The minimum and maximum differences between the DFT and IDFT electron densities are $-0.023$~$n_0$ and $0.026$~$n_0$, 
respectively, while the minimum and maximum differences between the electrostatic potentials are $-0.142$~$V_0$ and 
$0.003$~$V_0$, respectively. Here, $n_0$ and $V_0$ are the bulk average electron density and the bulk mean inner potential 
of  WSe$_2$, respectively \cite{footnote}. Positive values denote higher electron densities and higher potentials in the DFT method.

Electrostatic potentials simulated using the three methods were taken as input for the calculation of electron-optical phase images. 
The evolution of the real-space wavefunction of the electron beam in a TEM passing through the potentials was 
calculated using the multislice method \cite{cowley} implemented within the Dr. Probe software package \cite{barthel}. 
In this method, the sample is divided into a number of slices along the direction of the incident electron beam (the $z$~direction).
In the present calculations, the number of slices was chosen to be equal to the $z$-sampling in the DFT simulations, and,
hence, the potentials of the individual atoms were sub-sliced.

The following discussion is limited to the cases of
monolayer and bilayer WSe$_2$, in which the effect of
sample tilt on the measured average electron phase is
small and where dynamic scattering effects play
only a small role \cite{florian}.
In addition to an incident electron energy of $80$~keV, the parameters applied to the simulations included the aperture size,
the sample tilt, the Debye-Waller parameters for the treatment of damping effects due to thermal atomic vibrations,
as well as parameters for a quasi-coherent image wave convolution in order to take into account  
image vibrations and sample drift accumulated over
the long hologram exposure time of $12$~s. 
For the simulation of monolayer WSe$_2$, the parameters for the simulation were chosen according to typical experimental values.
In the case of bilayer WSe$_2$, the simulation parameters as well as the 
parameters for the correction of residual aberrations in the experimental phase
were determined from a Nelder-Mead minimization of the root mean square of the differences
between the 13 strongest beam amplitudes in the fast Fourier transform of the simulated and
experimental phase images \cite{nelder}. The experimental beam amplitudes were taken from a clean and almost defect-free area 
of the wavefunction presented in \cite{florian} that included 15 orthorhombic unit cells (Supplementary Material Fig. 1).
The empirically chosen parameters for the simulation of the WSe$_2$ monolayer, as well as the optimized parameters
for the simulation of the WSe$_2$ bilayer and the correction of the residual aberrations in the experimental bilayer
phase image, are listed in the Supplementary Material
together with an overview of the agreement achieved between the experimental and simulated phase images for bilayer WSe$_2$.

\begin{figure*}
\includegraphics{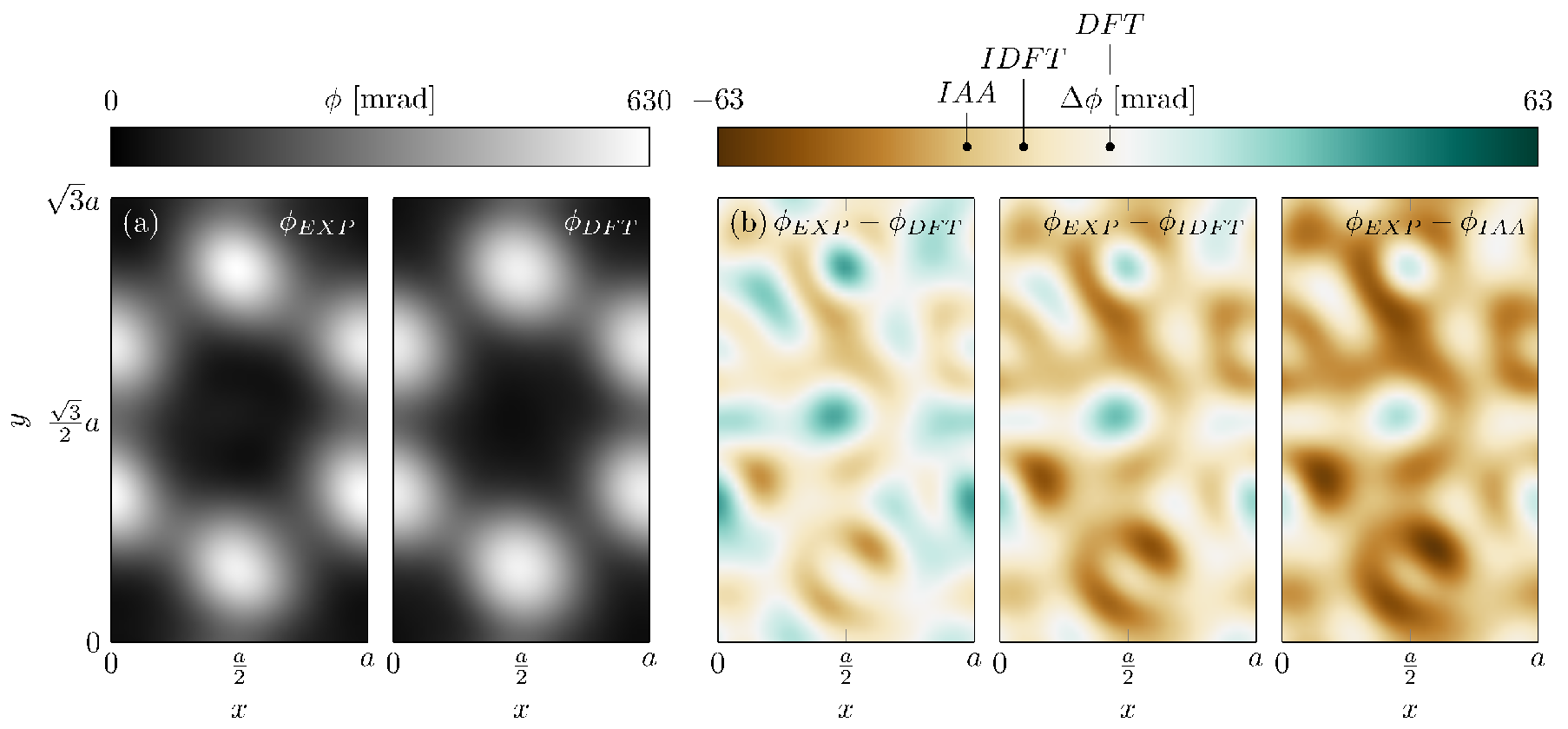}
\caption{\label{fig:phases}Comparison between simulated and experimental phase images of a WSe$_2$ bilayer structure, shown for a region corresponding to the area marked by a blue rectangle in Fig.~\ref{fig:geometry}. 
(a) shows an averaged experimental phase image of a WSe$_2$ bilayer \cite{florian}, alongside a
calculated phase image obtained by using the simulated electrostatic potential from the DFT method,
(b) shows differences between phase images obtained using simulated electrostatic potentials from the DFT, IDFT and IAA
methods and the experimental phase image.
Positive values correspond to higher phases in the experimental phase image. The markings on the color bar represent
the average differences for each of the simulation methods.
}
\end{figure*}

Spatially averaged phases obtained from the multislice calculations are listed 
alongside our experimental results in Table~\ref{tab:phases}. It should be noted that aberration correction has no influence on the averaged
electron phases presented here.
Corresponding simulated phase images for a 
WSe$_2$ bilayer 
are shown alongside a cell-average of the experimental phase image in Fig.~\ref{fig:phases}. 

For both monolayer and bilayer WSe$_2$, Table~\ref{tab:phases} shows that the DFT method yields the lowest average phase, while the IAA and IDFT results exceed the value obtained by the DFT method by approximately 9\% and 6\%, respectively. 
Remarkably good agreement is obtained between the average phase shifts obtained using the DFT-based simulations and the experimental values.
For the WSe$_2$ monolayer, the spatially-averaged phase obtained using the DFT method lies within the $1\sigma$ confidence interval for the extrapolated experimental value.
For the WSe$_2$ bilayer the discrepancy is larger but still within $1\sigma$ confidence interval.
In contrast, 
the values obtained from the IDFT and IAA methods deviate significantly more from the experimental values.

\begin{table}
\caption{\label{tab:phases}Spatially averaged electron phase shifts for WSe$_2$ monolayer and bilayer
structures obtained from electrostatic potentials obtained using the three different methods indicated. Experimental values are 
taken from wavefunctions presented in \cite{florian}.}
\begin{ruledtabular}
\begin{tabular}{lll}	
\hline
 &  Monolayer [mrad] & Bilayer [mrad]\\
 \hline
DFT &  \ \ \ \ \ \   $127.2$ &  \ \ \ \ \ \   $243.5$  \\
IDFT  & \ \ \ \  \  \  $134.5$ &  \ \ \ \ \ \  $256.7$  \\
IAA &   \ \ \ \ \ \  $138.4$ &  \ \ \ \ \ \  $265.4$  \\
Exp. &  \ \ \ \ \ \    $126\pm5$\footnote{The experimental spatially averaged phase for a monolayer is determined from an extrapolation of values acquired for thicker structures and several samples.} &  \ \ \ \ \ \  $240\pm10$  \\
\hline
\end{tabular}
\end{ruledtabular}
\end{table}

The experimental and calculated phase distributions in Fig.~\ref{fig:phases}~(a) show a good match. 
The notable elliptical distortion at the positions of the atomic cores can be attributed to anisotropic 
image shift fluctuations resulting from sample vibrations, drift or electrical instabilities of the lenses
during the hologram exposure time of $12$~s.
From Fig.~\ref{fig:phases}~(b), it is apparent that the differences between the experimental and simulated phase images are
mainly due to an offset in phase, which is smallest for the DFT method. The residual fluctuations of approximately
$13$~mrad are comparable to the vacuum phase noise of approximately $10$~mrad, while a comparison between
the phase images obtained with the DFT and IDFT methods (Supplementary Material Fig.~2) 
suggests that a noise level below $3$~mrad would be necessary in order to resolve the spatial signature of bonding.
The results, hence, indicate that,
although the change in spatial electron density resulting from the effects of bonding and the associated change in
electrostatic potential are non-homogeneous effects and are only likely to be measurable when phase images with much better
signal to noise ratios are available experimentally, the dominant effect of bonding on the measured mean electron phase is
accessible from the present results.

We also studied interlayer coupling in the DFT model, which does not include van der Waals forces and, hence, only
accounts for covalent effects. For this purpose, DFT and independent  layer DFT (ILDFT) results for a WSe$_2$ bilayer
were compared.
In ILDFT, the electrostatic potentials of the two layers forming a bilayer were calculated individually 
and then superimposed. 
The combined electrostatic potential was then used as input for a further multislice simulation, yielding 
an average electron phase shift of $243.6$~mrad, which differs by only $0.04\%$ from the DFT value for a WSe$_2$ 
bilayer (Tab.~\ref{tab:phases}). Since the main contribution in the interlayer coupling of two-dimensional materials
is given by van der Waals forces, this small effect of covalent interlayer coupling is reassuring.
Consequently, it is not expected that the electron beam will be sensitive to a shift in charge generated by
covalent interlayer coupling.
In order to estimate the effect of structural changes induced by
van der Waals forces, which are expected to modify the interlayer distance $w$,
we re-calculated the average electron phase in the DFT method for a bilayer with a decreased interlayer distance.
In our calculations, decreasing $w$ for a WSe$_2$ bilayer by $6\%$ ($0.2$~\AA) led to a decrease of the average 
electron phase shift by only $0.03\%$, confirming the weak effect of interlayer coupling.

In conclusion, the electrostatic potentials of few-layer WSe$_2$ structures have been calculated using progressively more accurate methods
and used as input for multisclice simulations of electron-optical phase images, for comparison with experimental results measured using off-axis electron holography.
Our results demonstrate that a perfect contrast match can be achieved between experimental and simulated 
atomic-resolution phase images for a sample of precisely know thickness.
Excellent agreement between simulated and experimental spatially averaged phase shifts is obtained when the effects of atomic bonding are taken into account in the simulations. 
If bonding effects are neglected, then the average phase can be overestimated in the simulations by up to 9\%\ 
for a WSe$_2$ monolayer. This overestimate of the electron phase results from a change in electrostatic 
potential associated with a small redistribution in electron density along the bonding directions between the crystal atoms. 
This conclusion was confirmed by comparing theoretical and experimental results.

We employed a fast and accurate combination of DFT calculations using pseudopotentials and all-electron atomic corrections to
restore core charge densities.
Due to the low computational cost of this approach, it should allow quantitative analyses of defects and substitutional
atoms in TMDs and other materials when large supercells are required, similar to high-resolution 
transmission electron microscopy studies on nitrogen-substitutions in graphene \cite{meyer}.

% In contrast to the large effect of intralayer bonding on the measured electrostatic potential, covalent interlayer bonding 
% in WSe$_2$ has a very small effect on the measured phase.
% As a further refinement to interlayer bonding, we suggest to include van der Waals forces and potentials in such calculations.

\section{Acknowledgments}

The authors gratefully acknowledge computing time granted by the John von Neumann Institute for Computing (NIC) and
provided on the supercomputer JURECA in the J\"ulich Supercomputing Centre (JSC) (JARA-HPC projects JIAS16 and JPGI90).
Z.Z. acknowledges financial support from the European Commission under the Marie-Curie fellowship (PIEF-Ga-2011-300036) and by the Deutsche
Forschungsgemeinschaft (DFG, German Research Foundation) grant ZA 780/3-1.
M.J.V. acknowledges a PDR grant from the Belgian Fonds National pour la Recherche Scientifique (GA T.1077.15) and ARC grand AIMED 15/19-09.
J.B. acknowledges funding from the German Science Foundation (DFG) grant MA 1280/40-1.
R.D.B acknowledges funding from the European Research Council under the European Union's Seventh Framework Programme (FP7/2007-2013)/ ERC grant agreement number 320832.

\pagebreak
\widetext

\begin{center}
\textbf{\large The Effect of Intra-Layer Bonding on Electron-Optical Phase~Images of~Few-Layer~WSe$_2$ - Supplementary material}
\end{center}

\begin{center}
\begin{figure*}[h!]
\includegraphics{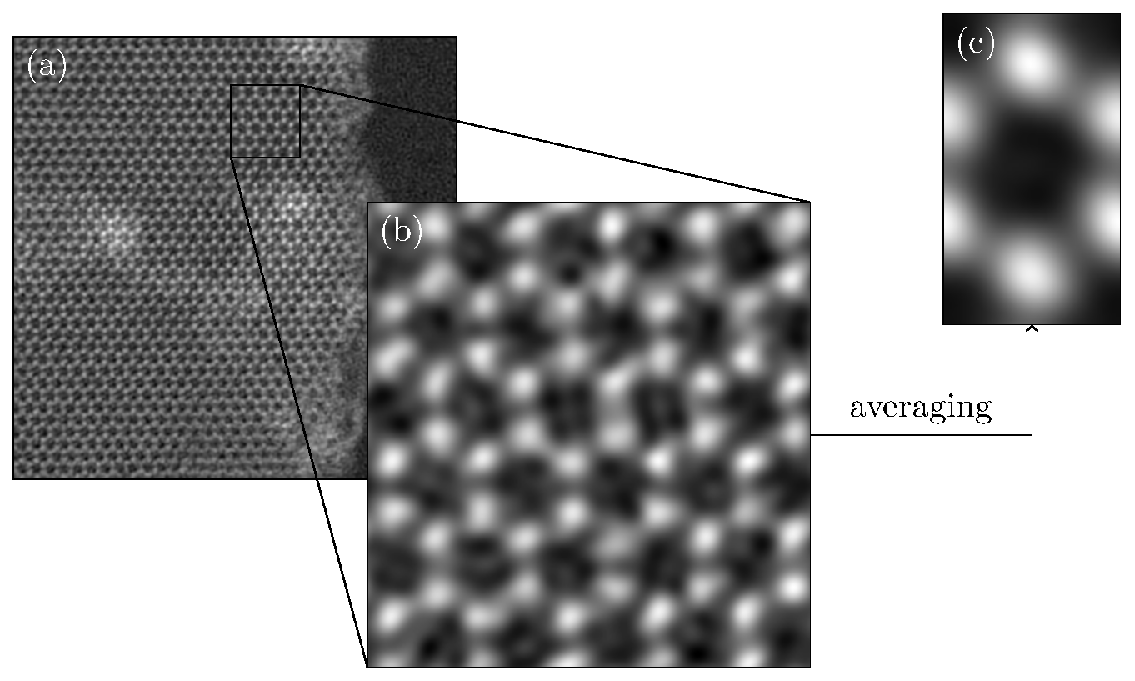}
\caption{\label{fig:phase}Aberration-corrected experimental phase image of a WSe$_2$ bilayer. (a) shows the full experimental phase image of the electron wave function
presented in \cite{florian}. (b) shows a zoomed-in view of an area that appears to be clean and almost free of defects. The area has
a size of 5x3 orthorhombic unit cells. (c) shows the cell-averaged phase image of the area shown in (b).}
\end{figure*}
\end{center}

\begin{table}[h!]
\begin{center}
\begin{tabular}{lcc}
 \hline
 & Monolayer & Bilayer \\
 \hline
 Electron energy & $ E = 80$~keV & $ E = 80$~keV  \\
 Aperture & $a=15$~mrad & $a=42$~mrad \\
 Debye-Waller factors & $B_{Se} = B_{W} = 0.3$~\AA$^2$ & $B_{Se} = B_{W} = 0.333$~\AA$^2$\\
 Specimen tilt & $t_x = t_y = 0^{\circ}$ & $t_x = -2.74^{\circ}$, $t_y = 0.41^{\circ}$ \\
 Image wave convolution & $s_1 = s_2 = 22$~pm & $s_1 = 43$~pm, $s_2 = 32$~pm, \\
 & & $s_1 \sphericalangle x = -56^{\circ}$\\
 \hline
 \end{tabular}
\caption{\label{tab:}Simulation parameters for the multisclice simulations of electron-optical phase images for WSe$_2$
mono- and bilayer. 
The simulation parameters for the WSe$_2$ monolayer were chosen according to typical experimental values, whereas
the simulation parameters for the
WSe$_2$ bilayer, excluding the electron energy and the aperture, were obtained together with the parameters for the correction of the residual aberrations
in the experimental phase image (Tab. \ref{tab:aberration}) from a Nelder-Mead minimization of the root mean square of the differences
between the 13 strongest beam amplitudes in the fast Fourier transform of both the simulated and
experimental phase images \cite{nelder}. }
\end{center}
 \end{table}

 \begin{table}[h!]
\begin{center}
$
\begin{array}{r c c r c c}
A_1 = & 1.08 \text{nm} & (-115^{\circ}) & C_1 = & -3.68 \text{nm} &\\
A_2 = & 174 \text{nm} & (-134^{\circ}) & B_2 = & 135 \text{nm} & (-96^{\circ}) \\
A_3 = & 2.29 \mu\text{m} & (-168^{\circ}) & S_3 = & 2.84 \mu\text{m} & (-135^{\circ}) \\
C_3 = & 13.6 \mu\text{m} &  & C_5 = & -6.5 \mu\text{m} &  \\
\end{array}
$
\caption{\label{tab:aberration}Parameters for the correction of the residual aberrations in the experimental phase image
of the WSe$_2$ bilayer (Fig. \ref{fig:phase}). The parameters were obtained together with the parameters for the simulation
of the WSe$_2$ bilayer from a Nelder-Mead minimization of the root mean square of the differences
between the 13 strongest beam amplitudes in the fast Fourier transform of the simulated and
experimental phase images \cite{nelder}.}
\end{center}
\end{table}

\begin{table}[h!]
\begin{center}
\begin{tabular}{lcccc}	
\hline
 &  Minimum Diff.  & Mean Diff. &  Maximum Diff. & RMS Diff. \\
 &   [mrad] &  [mrad] &    [mrad] & [mrad] \\
 \hline
IAA &   $-61$ &  $-25$ & $18$ & $28$ \\
IDFT &   $-52$ &  $-16$ & $29$ & $21$ \\
DFT &   $-36$ &  $-3$ & $38$ & $13$ \\
\hline
\end{tabular}
\caption{\label{tab:agreement}Agreement between the spatially resolved
experimental and simulated electron phases. The table shows the minimum, mean and
maximum difference in the electron phase, as well as the root mean square of the difference for the three different simulation
methods used in this work.}
\end{center}
\end{table}

\begin{center}
\begin{figure*}[h!]
\includegraphics{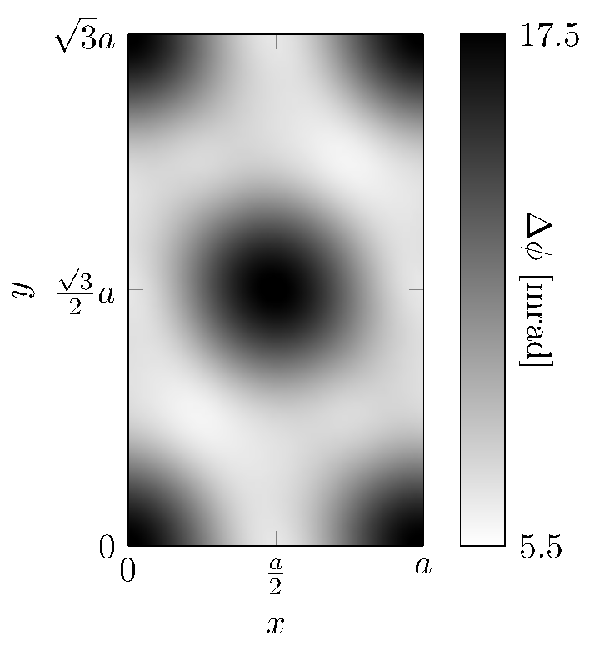}
\caption{\label{fig:phase_diff}Difference in the electron phase calculated for a WSe$_2$ bilayer using the DFT and IDFT methods.
The shown area 
corresponds to the area marked by a blue rectangle in Fig. 1. of the main text. Positive values denote 
a larger electron phase in the IDFT method. It is apparent that the largest differences in the electron phase can be found
along the bonding directions of the WSe$_2$ crystal whereas only small differences are found in the interstitial areas. The standard
deviation of the difference image is $3$~mrad.}
\end{figure*}
\end{center}
% \clearpage


\begin{thebibliography}{99}
\bibitem{footnote}
  Here, we approximate the mean inner potential of bulk WSe$_2$ with the average electrostatic potential of the central layer in the WSe$_2$ pentalayer structure.
\end{thebibliography}
\end{document}